\documentclass[information,article,accept,moreauthors,pdftex,10pt,a4paper]{mdpi}
\usepackage{array}
\firstpage{1}
\articlenumber{x}
\doinum{10.3390/------}
\pubvolume{xx}
\pubyear{2017}
\copyrightyear{2017}
\externaleditor{Academic Editor: David Bawden}
\history{Received: 23 May 2017; Accepted: 23 June 2017; Published: date}

\usepackage{soul}
\usepackage{upgreek}
\usepackage{enumitem}
\usepackage{subfigure}
\usepackage{booktabs}
\usepackage{multirow}
\usepackage{longtable}
\usepackage{microtype}
\makeatletter
\g@addto@macro{\UrlBreaks}{\UrlOrds}
\makeatother

\setitemize{parsep=6pt,itemsep=0pt,leftmargin=*,labelsep=5.5mm,align=parleft}
\setenumerate{parsep=6pt,itemsep=0pt,leftmargin=*,labelsep=5.5mm,align=parleft}
\setlist[description]{itemsep=0mm}

\Title{An Overview on Evaluating and Predicting Scholarly Article Impact}

\Author{Xiaomei Bai $^{1,2}$, Hui Liu $^{1,}$*, Fuli Zhang $^{3}$, Zhaolong Ning $^{1}$, Xiangjie Kong $^{1}$, Ivan Lee $^{4}$ and~Feng~Xia~$^{1}$}

\AuthorNames{Xiaomei Bai, Hui Liu, Fuli Zhang, Zhaolong Ning, Xiangjie Kong, Ivan Lee and Feng Xia}

\address{%
$^{1}$ \quad School of Software, Dalian University of Technology, Dalian 116620, China; xiaomeibai@outlook.com (X.B.); Zhaolongning@dlut.edu.cn (Z.N.); xjkong@ieee.org (X.K.); f.xia@ieee.org (F.X.)\\
$^{2}$ \quad Computing Center, Anshan Normal University, Anshan 114007, China\\
$^{3}$ \quad Library, Anshan Normal University, Anshan 114007, China; zfuli@outlook.com\\ % please add full labrary information
$^{4}$ \quad School of Information Technology and Mathematical Sciences, University of South Australia, SA 5095, Australia; ivan.lee@unisa.edu.au % please add zip code
}

\corres{Correspondence: liuhui1126@dlut.edu.cn; Tel.: +86-0411-6227-4391} % please to replace with a fixed tel number

\abstract{Scholarly article impact reflects the significance of academic output recognised by academic peers, and it often plays a crucial role in assessing the scientific achievements of researchers, teams, institutions and countries. It is also used for addressing various needs in the academic and scientific arena, such as recruitment decisions, promotions, and funding allocations. This article provides a comprehensive review of recent progresses related to article impact assessment and prediction. The~review starts by sharing some insight into the article impact research and outlines current research status. Some core methods and recent progress are presented to outline how article impact metrics and prediction have evolved to consider integrating multiple networks. Key techniques, including statistical analysis, machine learning, data mining and network science, are discussed. In particular, we highlight important applications of each technique in article impact research. Subsequently, we discuss the open issues and challenges of article impact research. At the same time, this review points out some important research directions, including article impact evaluation by considering Conflict of Interest, time and location information, various distributions of scholarly entities, and rising stars.}


\begin{document}

\section{Introduction}
Scholarly impact acts as one of the strongest currencies in the academia, and it is frequently measured in terms of citations of research articles. Citations indicate the impact of scholars, articles, journals, institutions, and other scholarly entities~\cite{aguinis2012scholarly}. The influence of an article is often quantified as an index, which represents its contributions for improving research finding by other scholars~\cite{gargouri2010self}.

\begin{figure}[H]
  \centering
  %\centerline{\epsfig{file=networks2.eps,width=\linewidth}}
  \includegraphics[width=0.6\linewidth]{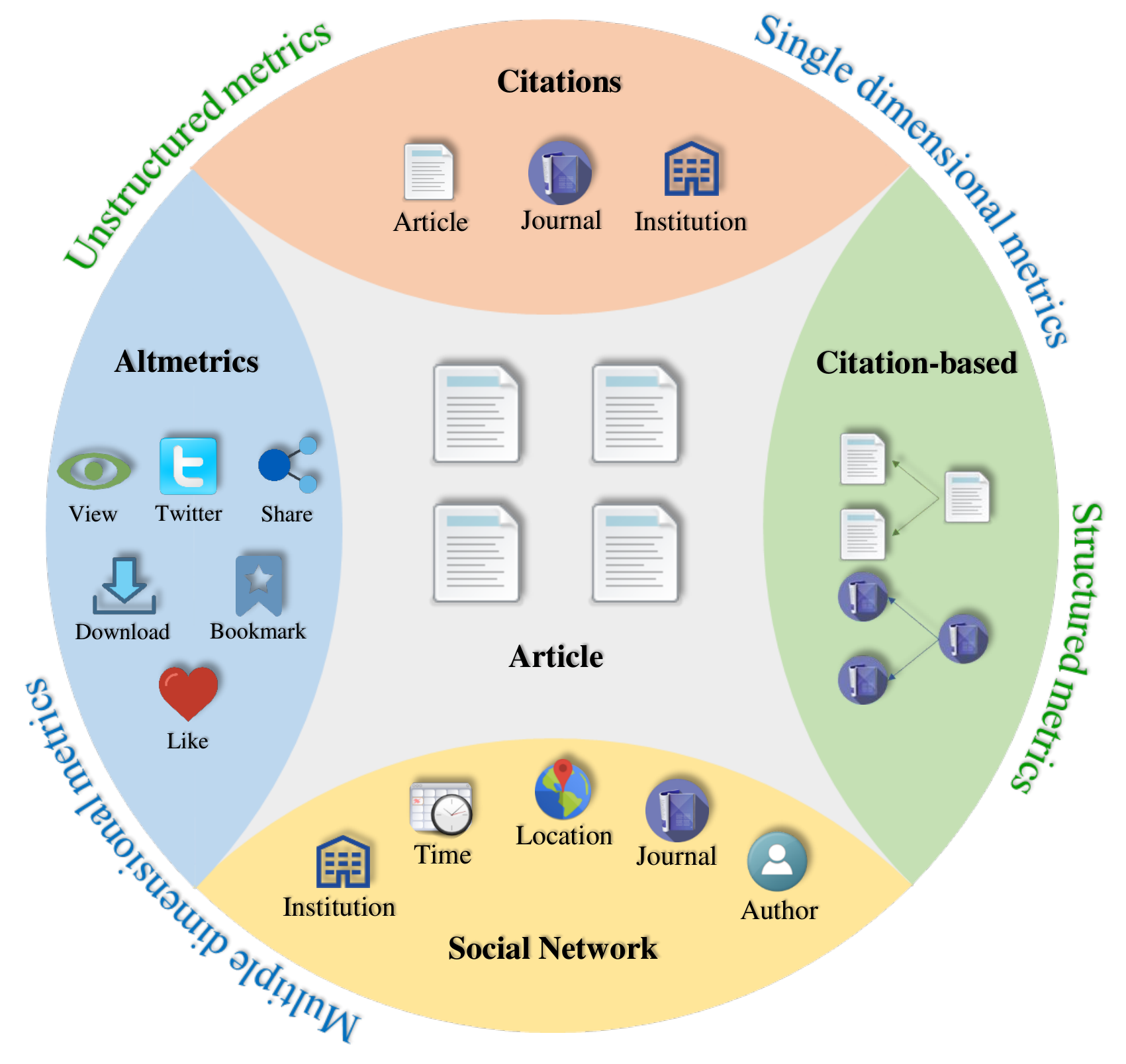}
  \caption{Methods of evaluating and predicting article impact.}
  \label{figure1}
\end{figure}

Researching the impact of scientific articles mainly focuses on two interrelated questions: how to assess the past impact of an article, and how to accurately predict its future impact? The study of article impact is important for evaluating the impact of individual scientists, journals, teams, institutions, and even for countries. It is also crucial for addressing the following fundamental problems, such as rewards, funding allocation, promotion, and recruitment decisions. Evaluating and predicting article impact have attracted great attention in the academic and scientific arena over the past decades. The~changes occur from one dimension to multiple dimensions, from unstructured metrics to structured metrics (Figure~\ref{figure1}). Citations~\cite{wang2013quantifying} are a popular indicator to measure article impact. However, it only focuses on the perspective of single dimension. Altmetrics~\cite{piwowar2013altmetrics} provide information on downloads, views, shares, and citations to assess article impact from a multidimensional perspective. PageRank has been introduced to evaluate article impact~\cite{chen2007finding}, which can be viewed as a milestone in impact research. It~has shown a structured method to quantify article impact. Meanwhile, in order to objectively evaluate article impact and accurately predict its future impact, machine learning and data mining techniques play crucial roles, such as mining the important characters of scholarly networks and optimizing the performance of algorithms~\cite{jordan2015machine}.

\begin{figure}[H]
  \centering
  %\centerline{\epsfig{file=networks2.eps,width=\linewidth}}
  \includegraphics[width=\linewidth]{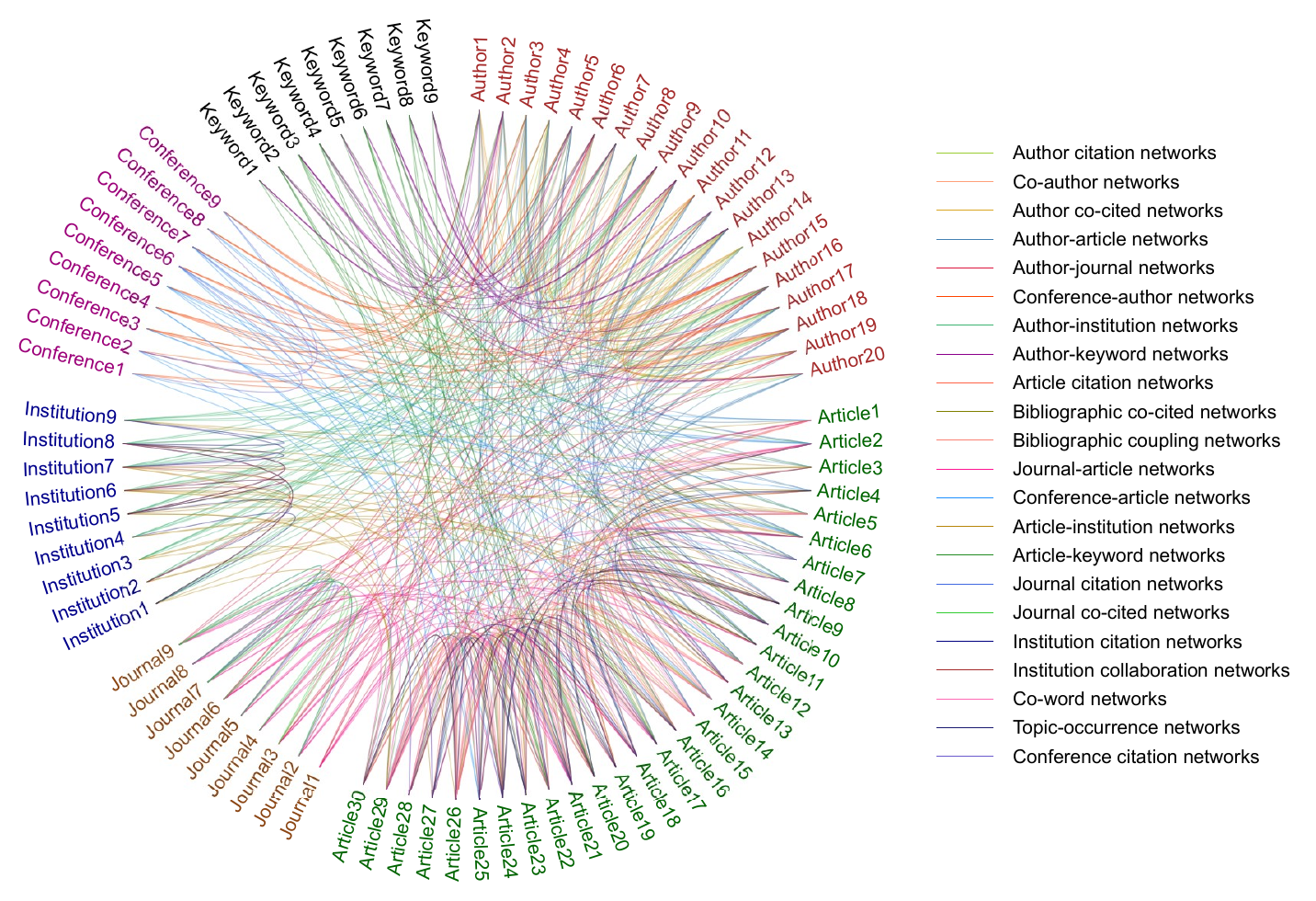}
  \caption{Characterizing scholarly networks.}
  \label{figure2}
\end{figure}

What drives the rapid development in evaluating and predicting article impact? The past decade has witnessed the rapid growth in the ability of network platforms to gather and transport a large number of academic data, i.e., a phenomenon usually referred to as ``Big Scholarly Data'' (see Figure~\ref{figure2}). Different networks with various scholarly entities and their relationships can be observed from Figure~\ref{figure2}. Scholars can collect such data to solve the problems of scholarly impact evaluation. They~can obtain useful insights from such datasets by leveraging statistical analysis, machine learning, data mining, and~network science techniques. The academic data with exponential growth become essential to develop the scholarly impact metrics. The metrics combine the statistical and computational considerations. However, one problem cannot be ignored. That is these datasets are personalized. SCOPUS contains abstracts and citations of journal papers. Web of Science offers online scientific citations by Thomson Reuters. PubMed includes more than 23 million citations for biomedical literature. CiteULike allows users to search and share scholarly papers. Mendeley can not only be used to manage references, but also it is an academic social platform. Digital Bibliography $\&$ Library Project (DBLP) shows publications of journals and conferences, not including citation information. Microsoft Academic Graph (MAG) includes heterogeneous information with publication records, authors, institutions, journals, conferences, fields of study and citation relationships. In these raw data, the most prominent problems are loss and incompletion of data, which probably will result in poor performance of evaluation and prediction to some extent. Data cleaning and supplement are necessary for accurately capturing the evaluative and predictive results. Besides, these data sets can be jointly investigated to complement one another. For example, DBLP does not include citation information, but it has an effective mechanism to process name disambiguation. Integrating DBLP dataset and the citation information of SCOPUS can meet the needs for some scholarly analysis.

\section{Key Techniques}
In this section, we discuss four crucial techniques for evaluating and predicting article impact including statistical analysis, machine learning, data mining and network science techniques.
%~\cite{Radicchi2008Universality}.
\subsection{Statistical Methods}
Statistical methods cover the process of collecting, dealing with, analysing and explaining data. Researchers can gain science knowledge from data through statistics. Statistical analysis is mainly interested in analyzing and understanding data, including regression models, variable selection, principal component analysis, factor analysis, cluster analysis, canonical correlation analysis, time series analysis, probability and density estimation, and so on~\cite{johnson2002applied,di2012advanced}. Regression models contain single variable regression and multiple variables regression. Statistical techniques are usually used in most fields of nature and social science, such as finance, medical treatment, industry etc. In researching scholarly impact fields the benefits of statistical techniques are as follows:
\begin{itemize}[leftmargin=*,labelsep=5mm]
  \item Pre-process data;
  \item Optimize parameters, for instance, multiple variables linear regression;
  \item Select features and improve models for scholarly evaluation and prediction. For example, use the massive existing statistics to estimate a probability density function;
  \item Analyse scholarly data to obtain statistical data, and then use statistical model to predict the trends of impact, top scholars, top articles, etc.
\end{itemize}

To explore the relationships between citations and citation distance, statistical analysis such as grouping and clustering may be applied. When we use group analysis technique, an appropriate segmentation point of citation distance is crucial. Usually, the selection of segmentation point depends on the experimental data. An advantage of group analysis technique is relatively easy to deal with data. However, due to the compulsory group, the disadvantage of group analysis is obvious. Clustering analysis technique remedies the drawback of compulsory group such as Density-based Spatial Clustering of Applications with Noise algorithm can be used to analyze the relationships between citations and citation distance based on the density of institutions.

In short, statistical analysis can be used to pre-process data, capture intermediate results or gain final results in evaluating and predicting research of scholarly impact. For example, a multivariate linear regression was used to estimate the parameters of three algorithms for evaluating the impact of papers~\cite{bai2016identifying}. Based on principal component analysis, a factor analysis was used to explore the main components in bibliometric and altmetric indicators~\cite{costas2015altmetrics}. Especially when scholars predict citations of a paper or a scholar's H-index, they usually give an estimative range instead of a specific value.

\subsection{Machine Learning}
Machine learning is one of the most rapidly developing techniques, and it can help computers to address the problems learnt through experience~\cite{jordan2015machine}. Machine learning mainly includes three major paradigms: supervised learning, unsupervised learning, and reinforcement learning. Supervised~learning is widely applied in spam classifying of e-mails, face identifying, and medical diagnosis fields. It aims to generate predictions according to its mapping functions. Relying on different mapping functions, learning algorithms are divided into neural networks~\cite{rojas2013neural}, support vector machines (SVM)~\cite{hearst1998support}, decision trees~\cite{quinlan1986induction}, logistic regression~\cite{hosmer2004applied}, and decision forests~\cite{ho1995random}.  The mapping functions are driven by different kinds of application needs. Unsupervised leaning focuses on direct inference of predictions without the help of the training sample of previous solved cases~\cite{hastie2009unsupervised}. The purpose of reinforcement learning is to learn a mapping function by desponding on intermediate between supervised and unsupervised leaning in training data. Reinforce learning has been successfully applied in human-level control~\cite{mnih2015human}.

In recent years, one prominent progress in supervised learning involves deep neural networks. Deep learning~\cite{lecun2015deep} has played an important role in computer vision, speech recognition, natural language translation, and collaborative filter. Deep learning algorithms can be used to discover useful representations of the input data without the requirement of labelled training data. The development of machine learning is closely related to other research fields progress. As machine learning theory develops, we will see the benefits it brings us. Machine learning contributes to scholarly impact research as follows:
\begin{itemize}[leftmargin=*,labelsep=5mm]
  \item Design effective algorithms fitting to various scholarly sources of data;
  \item Predict future trends such as articles impact and scholars' impact in future;
  \item Conduct scholarly recommendation such as recommending collaborators, the articles with top impact in various research fields.
\end{itemize}

Currently, some researchers have leveraged machine learning techniques to successfully predict scholarly impact, including articles, scholars, institutions and even countries. The commonly used methods for predicting scholarly impact include neural networks, SVM, Markov~\cite{jarrow1997markov}, XGboost~\cite{chen2015xgboost}, etc. In term of the performance of predicting the scholarly impact, neural networks model is better than Markov model. Neural networks model can be used to deal with large amount of data compared to Markov model and SVM model. SVM model can not only be directly used to regress, but also be used to classify. The SVM model is more suitable to deal with a small amount of scholarly data. Otherwise,~mixing SVM model and neural networks model can obtain better performance of prediction compared to the independent SVM or the single neural networks model. The predictive power of XGboost is better than Markov, neural networks, and SVM. However, a disadvantage of XGboost is that it needs to adjust a large of number of parameters. In the future, we believe that machine learning can provide more support to resolve emerging issues about scholarly impact, and also can provide models for understanding learning in scholarly impact, biological evolution, neural systems and other research fields.

\subsection{Data Mining}
Data mining is used to discover knowledge hidden in a large amount of data, including spatial data mining, temporal data mining, sequence data mining and intention data mining~\cite{bhise2013importance}. Data~mining has important applications in finance, telecommunication, science, and engineering fields. In~recent years, we have witnessed a rapid expansion in the ability to collect data from various sensors and online media platforms in different formats. For instance, a large source of data is going to be generated from online platforms like Facebook, Twitter, and Google. Big data drives scholars to continually explore useful patterns for better services. Meanwhile, it gives a challenge for big data mining. Data~mining can benefit scholarly impact research as follows:
\begin{itemize}[leftmargin=*,labelsep=5mm]
  \item Mining heterogeneous academic networks, such as article-author networks, author-journal networks, author-institution networks, etc.;

  \item Exploring the complex relationships among academic entities, including the relationships of papers, authors, journals, conferences, teams, institutions and countries;
  \item Seeking automatically patterns in scholarly data to predict future trends and improve predicting~performance;
  \item Mining large data streams for effective scholarly recommendations;
  \item Cleaning scholarly data to gain valuable information;
  \item Integrating diverse kinds of scholarly data.
\end{itemize}

In brief, data mining can solve the scholarly evaluation and predication problems by analyzing data in database. Discovering meaningful patterns in scholarly networks will lead to some advantages. Useful patterns allow scholars to predicate scholarly impact based on new data. For example, in order to predict the impact of an article, we may first train the data of previous years by applying machine learning techniques, like neural network, Markov and SVM models. In addition, we predict future impact of an institution on testing datasets. How to express a pattern is important. The expressions of a pattern can be presented in two ways: transparent box and black box. The former's construction discloses the structure of the pattern by explaining something about scholarly data, while the latter's construction is inexplicable. Data mining also involves learning for finding structural patterns in scholarly data. It~helps to explain data before making predictions. In data mining, machine learning is applied in many research fields. It is used to capture the explicit knowledge structures which are important to preform well on new data.

\subsection{Network Science}
Network science can help to understand the structure of networks, development and weaknesses. Despite apparent diversities, a lot of networks generate, evolve, and are driven by some basic laws and mechanisms. For instance, degree distribution has been proved to be the power law; small world property is an important principle in many networks. Two important organizing principles of the evolution of networks were introduced, i.e. preferential attachment and fitness~\cite{barabasi2013network}.

\begin{figure}[H]
  \centering
  %\centerline{\epsfig{file=networks2.eps,width=\linewidth}}
  \includegraphics[width=0.7\linewidth]{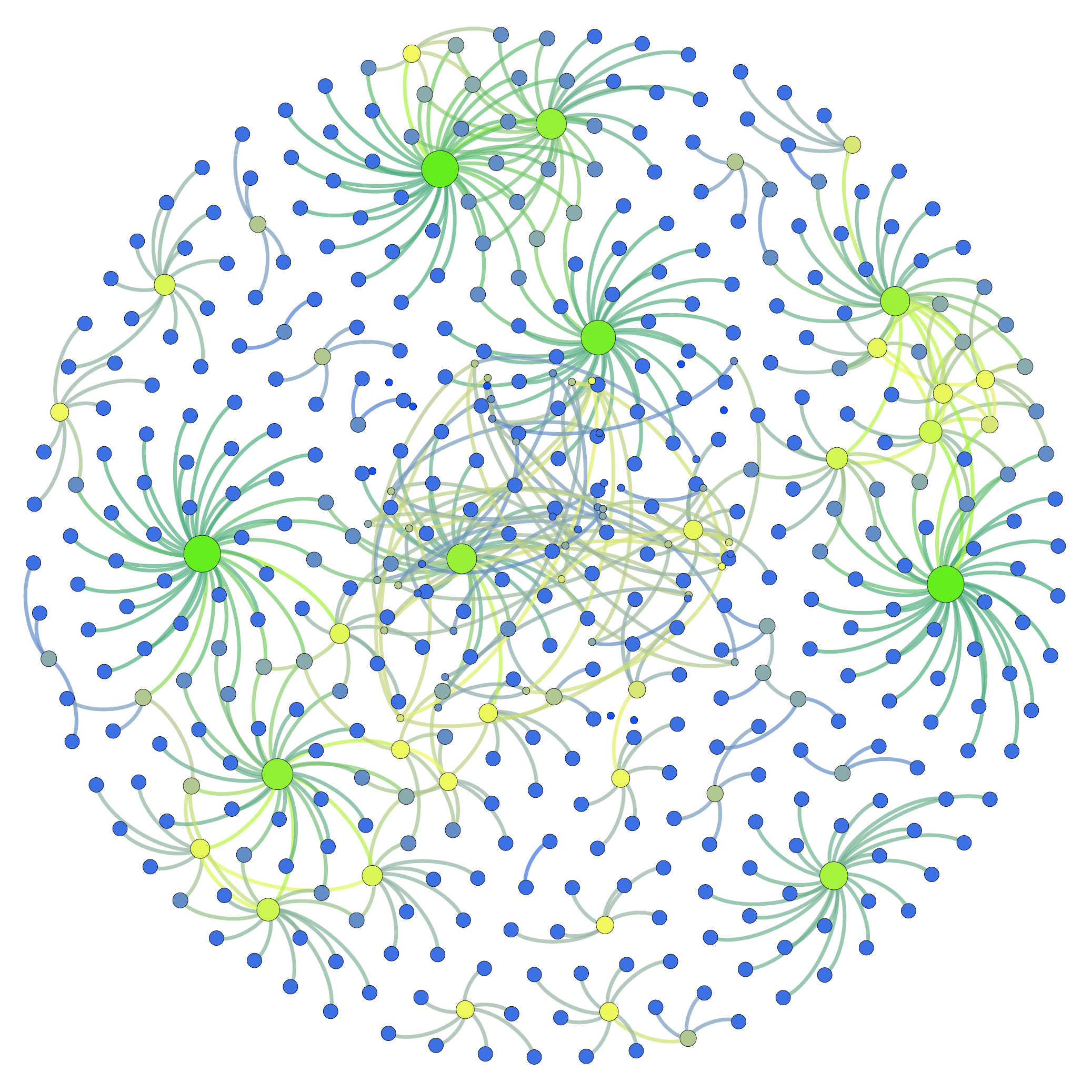}
  \caption{Characterizing citation relationships of article-article networks. The degrees of nodes range from small (blue) to large (green). The larger the degree of a node is, the more references the article has.}
  \label{figure3}
\end{figure}

Well-known scholarly network structures are complex, including homogeneous and heterogeneous networks~\cite{yan2012scholarly}, directed and undirected networks~\cite{west2013author}. Homogeneous citation networks contain article-article networks, author-author networks, journal-journal networks, and word-word networks, etc. Figure~\ref{figure3} shows the citation relationships of article-article networks generated by random extracted $486$ articles and their references from APS dataset, $561$ edges in total. Each circle represents an article and the links represent citation relationships. Blue, yellow and green represent nodes with small, medium and large degrees. Heterogeneous citation networks include article-author networks, author-journal networks, article-journal networks, etc. In particular, co-author networks and co-word networks are also important homogeneous networks in scholarly impact studies. Citation network is a~representative directed network, showing a link from a citing paper to a cited paper, while co-author network is a undirected network. The important indices of nodes in undirect networks include degree centrality, betweens centrality, closeness centrality, k-shell, k-core, and eigenvector centrality. In~directed networks, two representative algorithms, i.e., PageRank~\cite{page1999pagerank} and HITS~\cite{kleinberg1999authoritative} algorithms, are~commonly used to calculate the importance degree of nodes.

In diversified scholarly networks, network analysis plays a key role mainly in the following three aspects. First, network analysis helps identifying key nodes in scholarly networks. These key nodes are a series of scholarly entities, including top influential articles, top influential authors, top influential journals, top influential teams, top influential institutions, co-authors with super tie, academic rising star~\cite{zhang2016identifying,zhang2016rising}, serendipity in scientific collaboration~\cite{sugiyama2011serendipitous}, Sleeping Beauties in science~\cite{ke2015defining}, etc. We also need to study the difference of the important degree of various nodes in unweighted and weighted networks. Most scholarly networks are weighted~\cite{bai2016pncoirank,zhu2015measuring}, but we cannot always obtain appropriate weights. However, an appropriate weight is the key for quantifying scholarly impact. For example, impact of an article is no longer a simple citation count. The importance degree of each article in citation networks should consider the authors' authorities of citing articles and published journal's prestige of the article through analyzing the citation networks. Citation-based structured measurements have provided new perspective for evaluating scholarly impact. Second, network analysis helps to explore the most important structure features, such as what features determine scholars' success~\cite{sutherland2015constructions}, success of an article~\cite{letchford2015advantage}, and success of teams~\cite{anicich2015hierarchical}. Third, network analysis helps quantifying the relationships among scholarly entities, including articles, authors, journals, conferences, institutions, teams and countries. For example, previous researchers have quantified the relationships of co-authors in scientific community. It indicates scientific collaboration with weak, strong, and super ties from longitudinal perspective~\cite{petersen2015quantifying}. All in all, network structured analysis provides a solution to quantifying the scholarly impact.

In the next section, we will introduce article impact metrics and prediction, mainly including two aspects: core methods and their recent research progress.

\section{Article Impact Metrics}
\begin{figure}[H]
  \centering
  %\centerline{\epsfig{file=networks2.eps,width=\linewidth}}
  \includegraphics[width=0.9\linewidth]{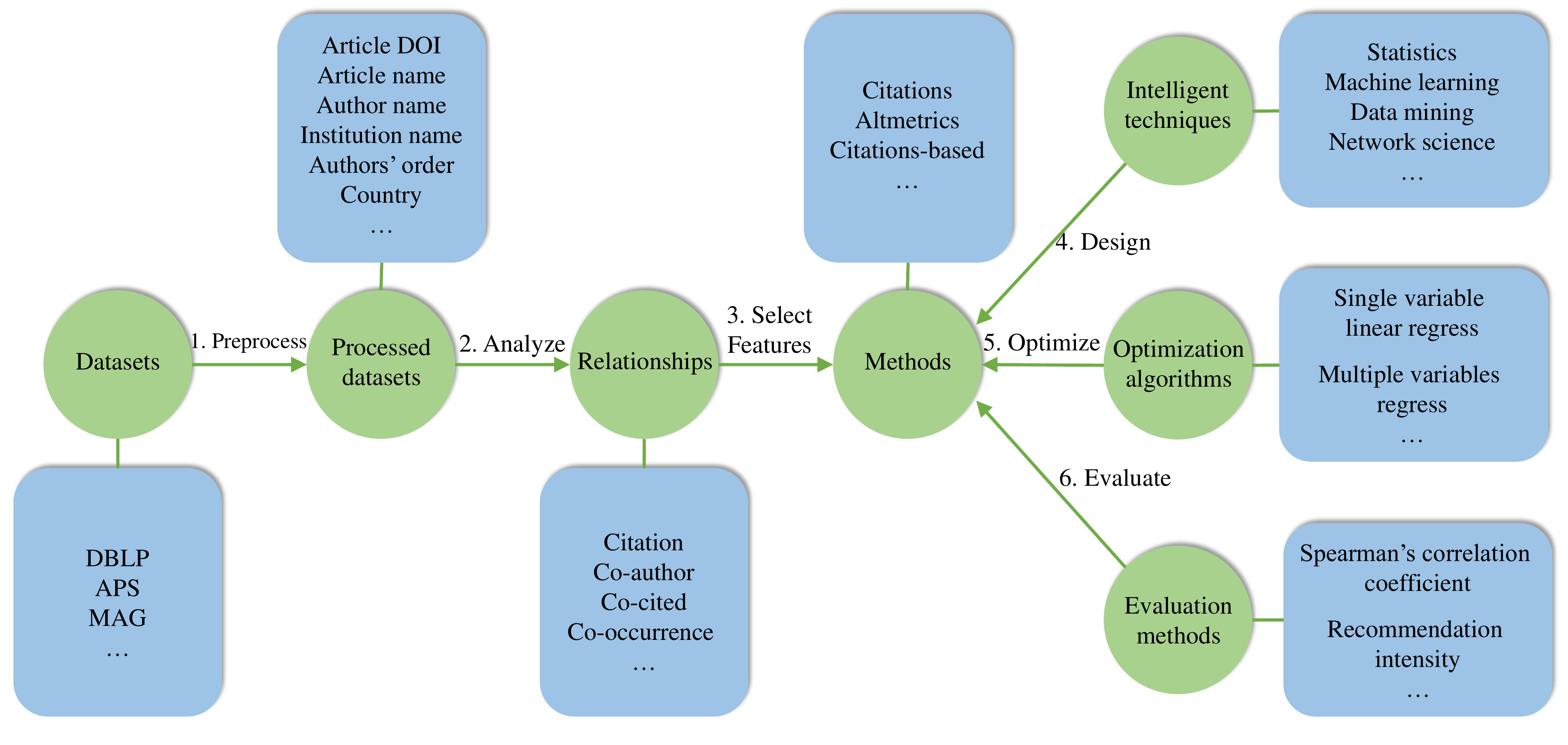}
  \caption{Frameworks of evaluating article impact.}
  \label{figure4}
\end{figure}

Figure~\ref{figure4} provides a framework for evaluation of article impact to build and test a set of scholarly data models, including data collection, data pre-processing, data analysis, features selection, algorithms design, optimizing algorithms and evaluation of algorithms. Datasets refer to original datasets like DBLP, APS and MAG. According to targets of evaluation, we can use the original datasets or complement the original datasets by crawling necessary data from scholarly websites like SCOPUS. Researching the various relationships including citation, co-author, co-cited, etc. is beneficial to assess the impact of scholarly impact. For example, identifying different citation relationships provides an objective evaluation method~\cite{bai2016identifying}. In the assessment framework, the assessment method is the most central part. Currently, there are several types of assessment methods: citations, Altmetrics and citations-based structured metrics. The validity of the verification method is an essential part. Common evaluation methods include Spearman's correlation coefficient, recommendation intensity and so on. Impact metrics can briefly be divided into two categories: unstructured metrics (or statistical metrics) and structured metrics according to the way of measurement.

Citations as statistical method are perhaps the oldest and most widely used metric for article impact evaluation. Citations as measuring metric are always under dispute. From the perspective of objective evaluation, can original citations truly characterize the quality of article? The answer is obviously no. The biggest obstacles are self-citation and mandatory citation~\cite{esfe2015mandatory}, which have increased the difficulty of objectively measuring article impact. How to accurately identify a variety of self-citation and mandatory citation is challenging. Meanwhile, negative citation has attracted scholars' attention~\cite{catalini2015incidence}. However, scientific researchers do not stay at distinguishing the citation patterns. The~scholarly publications are undergoing the changing from traditional prints to online platforms. The~change generates some open issues. Meanwhile, it presents an opportunity to characterize article impact from multidimensional perspective.

Altmetrics~\cite{priem2014altmetrics} emerge at the historic moment and obtains much attention in academic community. Altmetrics are the study of measuring the scholarly impact based on activities in social media platforms, and go beyond citations~\cite{kwok2013research}. Altmetrics present various quantitative values including citations, downloads, mentions, tweets, shares, views, discussions, saves and bookmarks from statistical perspective. Altmetrics scores (mentioned in blogs) can be used to identify highly cited articles. At the same time, Altmetrics can complement and improve evaluation of article impact with new insights ~\cite{costas2015altmetrics}. Although broadening the evaluation methods for measuring scholarly impact, Altmetrics lack the authority and credibility as metrics. It is partly because Almetrics are easy to be gamed by malicious scholars~\cite{cheung2013altmetrics}.

Citations-based structured methods have made some progress. One significant measurement of impact metrics in recent years involves homogeneous and heterogeneous networks~\cite{yan2010measuring}, including citation networks, co-author networks, co-citation networks, article-author networks, article-journal networks, author-journal networks. The diversity of scholarly networks can satisfy the diverse needs of applications with different scholarly structures capturing different kinds of scholarly characters. One~thing can be certain: citations-based structural metrics can generate a truer measure of the importance of an article than citations alone. Previous researchers have contributed to the structural metrics for evaluating article impact~\cite{chen2007finding,wang2013ranking,walker2007ranking,sayyadi2009futurerank,zhou2012quantifying,wang2014future,liu2014tri}. These assessment methods mostly are based on PageRank algorithm and HITS algorithm. PageRank algorithm provides a fast and objective ranking way to rank the nodes in network. In a citation network, papers with higher PageRank scores have more chances to be visited. PageRank is more suitable for homogeneous networks. In scholarly networks, HITS algorithm distinguishes the scholarly entities as authorities and hubs based on the local structure, and calculates their scores in a mutual reinforcing way. HITS algorithm can also be applied to heterogeneous networks like paper-author network and paper-journal network, in which the authors and journals are regarded as hub nodes, and the papers are regarded as authority nodes. It~is worth mentioning that S-index metric measured article impact through influence propagation in heterogenous citation networks~\cite{shah2015s}. Meanwhile, Neil Shah et al. suggested a good impact metric should consider the following six aspects: volume sensitivity, prestige sensitivity, robustness, extensibility, temporality, interpretability and computability. Exploiting network structure characters may provide an opportunity to develop a refined and objective metric for measuring the scholarly~impact.

While co-citation analysis can be utilised to associate the relevance across different disciplines and to identify the bridging nodes~\cite{small2010maps}, it should be noted that citation-based metrics are biased by diverse domain sizes and citation activities~\cite{kaur2013universality}. Domain variation may hamper a fair evaluation for scholarly impact, such as scholarly papers in some disciplines are cited much more or much less compared to others~\cite{radicchi2008universality}. Two important reasons cause the above results. One is uneven number of cited papers each article in different domains, the other is unbalanced cross-discipline citations. Although scholarly papers can be cited by different domains, Schneider et al.~\cite{schneider2017feasibility} suggested relative citation pattern within disciplines should be considered for the evaluation of scholarly impact.

\section{Article Impact Prediction}
\begin{figure}[H]
  \centering
  %\centerline{\epsfig{file=networks2.eps,width=\linewidth}}
  \includegraphics[width=0.55\linewidth]{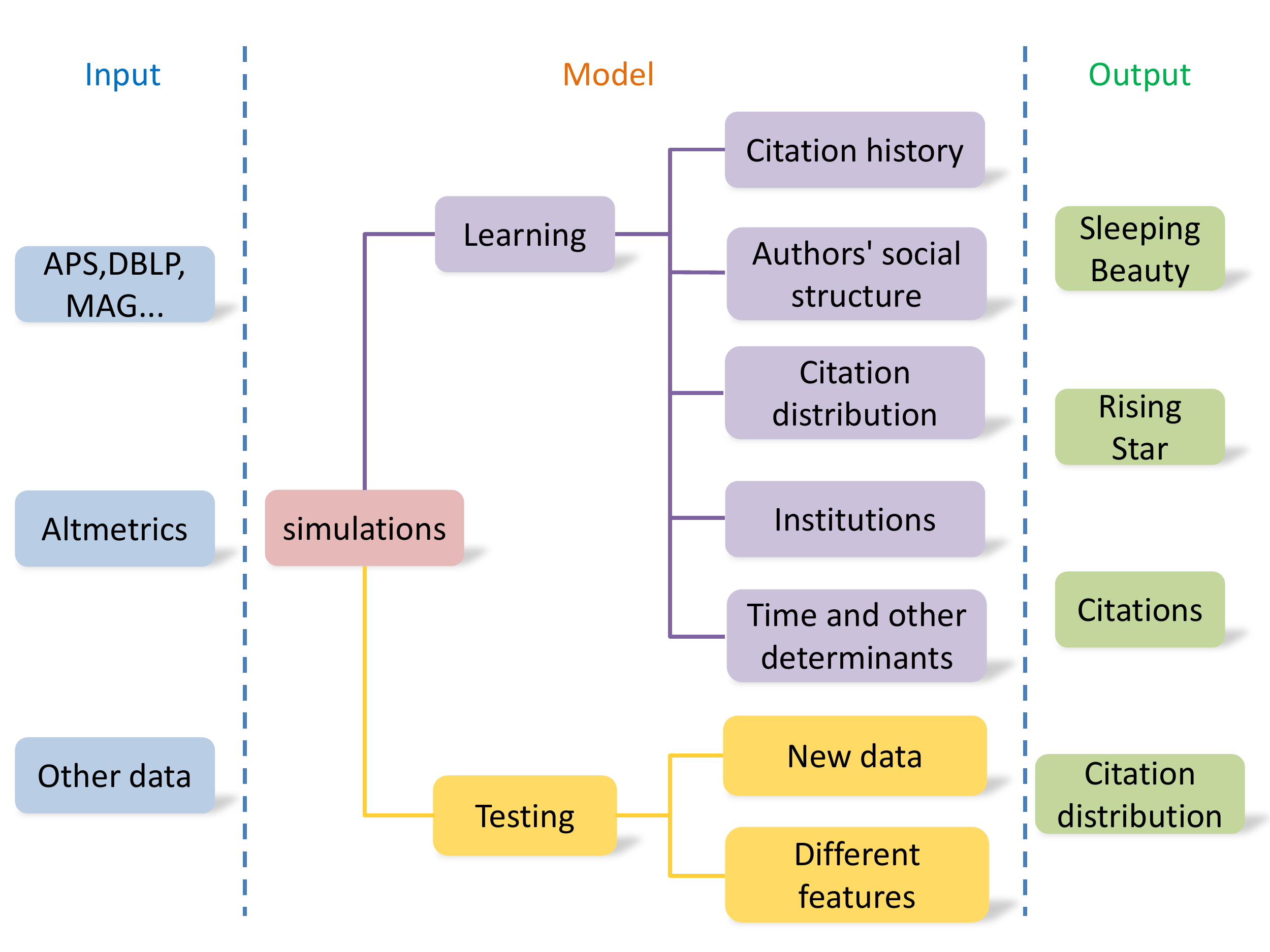}
  \caption{Flowchart of predicting article impact.}
  \label{figure5}
\end{figure}

Prediction of future impact is an emerging area, researching on the ``science of science''. Impact~prediction is more important compared to impact evaluation. Impact prediction can directly allocate funds, scientific awards, and other decisions. Figure~\ref{figure5} provides a flowchart of a computational model for predicting article impact. The left column (Input) is the input data, capturing publication, citation, downloads, reviews, and other information. The center column (Model) describes model learning and testing. The right column (Output) provides a few specific examples which the model can predict.

Specially, article impact prediction has attracted a lot of attention in recent years. Predicting~an article impact mainly focuses on predicting citations or citation distributions through network science, data mining and machine learning techniques (see Table~\ref{tab:1}). Early citations of an article played a critical role for predicting its long-run citation~\cite{bruns2015research}. They showed that university ranking with cumulative citations can be easily predicted by early received citations across the economics discipline at a university. Cao et al.~\cite{Cao2016A} presented a Gaussian mixture model to predict future citations of papers based on short-term citation activities. Peter et al.~\cite{Klimek2016Successful} constructed a keyword-term network to predict the numbers of citations in the future by analyzing the recursive centrality measures, indicating document centrality has higher predictive ability for the future citations of papers. Based on quantile regression, Stegehuis et al.~\cite{stegehuis2015predicting} proposed a model to predict the probability distribution for future citations of an article, and considered two key features: early citations and journal impact factor. Yu~et~al.~\cite{yu2014citation} leveraged four categories of features, including articles, authors, citations, and journals to predict future citations of an article based on stepwise regression analysis. Based on co-authorship networks, a Machine Learning Classifier was developed to predict whether a publication would get high citations~\cite{sarigol2014predicting}. Based on Random forest classifier, they showed a supervised classification model, in~which multidimensional feature vectors were considered to predict the future citations of a paper. Wang et al.~\cite{wang2013quantifying} constructed a generative model for predicting long-term impact of an article by using three key factors: preferential attachment, citation trend, and fitness. In short, previous researchers are mostly based on early citations for predicting the impact of paper. They mainly focus on the autocorrelation of historical data in citation network. However, a common drawback of these predictive methods is that they are dependent too much on historical citations. Exploring the fundamental characteristics of citations yielded may be able to find a novel predictive method, ignoring the early citations. In recent years, with the development of social media, social media activities are used to reflect the underlying impact of an article. For example, Tweets can predict whether an article can be cited frequently when an article was published for 3 days~\cite{eysenbach2011can}. Based on a heterogeneous scholarly network, Mohan et al.~\cite{timilsina2016towards} predicted academic impact by integrating the bibliometric data with the social data like weblogs and mainstream news, indicating that graph-based measure can reasonably predict the impact of early stage researchers.

\unskip
\begin{table}[H]
\centering
\caption{Several representative methods for predicting article impact.}
\begin{tabular}{ccc}
\toprule
%\begin{tabular}{@{\vrule widthpt}lrcccc}}
%\begin{tabular}{|c|c|c|}                                                       \hline
\textbf{Features}&\textbf{Prediction Goal}&\textbf{Main Techniques}\\ \midrule
early citations, Journal   &\multirow{2}{*}{quantile of citations distribution}& \multirow{2}{*}{quantile regression}\\
Impact Factor &&\\\midrule

%long-term citations (10 years) &   next year citations &   0.96013289\\ \hline
authors characteristics, & \multirow{4}{*}{citations}& \multirow{4}{*}{multivariate analysis}\\
 institutional factors, features &&\\
of article organization, &&\\
research approach&&\\\midrule

Social dimension: &\multirow{2}{*}{citations}& \multirow{2}{*}{random forest classifier}\\
co-authorship networks&&\\\midrule

year, page count, author count,    &\multirow{4}{*}{long-term citations} &\multirow{4}{*}{random forest}\\
author name, journal, abstract&&\\
length, title length, special&&\\
issue, etc.&&\\\midrule
\multirow{2}{*}{Altmetrics: tweeter}&\multirow{2}{*}{citations}   &correlation analysis, linear \\
&&regression analysis\\\bottomrule
\end{tabular}
\label{tab:1}
\end{table}
There is an increasing interesting in identifying Sleeping Beauties in science. Sleeping Beauty in scientific community refers to that the value of an article can be recognized only after years of publication~\cite{ke2015defining}. Ke et al. suggested a common mechanism using a parameter-free method to identify Sleeping Beauties on large-scale datasets.

\section{Open Issues and Challenges on Article Impact Metrics }
Despite pioneers have obtained success, article impact remains a young field with many open issues. In previous researches, many different datasets are usually used to quantify scholarly article impact. These granulitic and inconsistent data have been applied in various scholarly researches. Sharing datasets are necessary and valuable for objectively evaluating article impact and generating new metrics. Unified and consistent scholarly datasets are an open issue. Citation-based structured metrics are relatively new and have got less attention. Researchers consider that the important degrees of citation structures is newly shaped by PageRank and HITS algorithms introduced in scholarly networks. In addition, social dimensioned assessment and citation distributions have been less explored. Thus, multidimensional metrics for quantifying article impact are an open issue. Altmetrics~have been considered for complementing article-level metrics. Pioneered researchers have made some progress. Altmetrics for evaluating scholarly article impact is still an open issue. In this open issue, locating the reasonable and available benchmarks is an urgent need to be solved.
\subsection{Unified and Consistent Scholarly Datasets}
With the rapid emergence of a large number of social platforms, scholarly datasets present hitherto unknown event in academia. Even though these datasets possess personalized characters, they have the problems of missing data, repeated data, data uncertainty phenomena. Evaluation metrics based on these inconsistent datasets can bring some problems. For example, reproducing scientific findings in previous researches can be realized. Therefore, unified and consistent scholarly datasets should be ascertained and shared by scientific researchers in academia for impact metrics.
\subsection{Multidimensional Metrics}
In previous researches, citation-based structured metrics mainly consider the dimensions of authors, journals, articles and time. Each author's importance in citation networks is usually ignored. An article generative impact is regarded as the same no matter who cites it. In fact, citing authors' impact in citation networks should be investigated for objectively quantifying article impact. Copying the same citations from other articles is a frequently observed practice in academic publications~\cite{simkin2003read}. Therefore, an article may get more citations through frequency-dependent copying if it is cited by experienced scholars. The article impact can be influenced by many factors such as authors' social relationships, citation distributions of authors, journals, institutions and countries. In particular, identifying anomalous citation patterns and weakening citation strength are critical for objectively measuring article impact~\cite{bai2016identifying}. Although analysing Conflict of Interest (COI) relationships between authors has given a solution to identify anomalous citations. We need to mine COI relationships for more objective assessment in a further step. These problems have not been addressed. Therefore,~future impact metrics need to explore the importance in citation networks, authors' social relationships, various citation distributions, etc.
\subsection{Altmetrics}
Altmetrics are recent article-level metrics~\cite{thelwall2016data}. Altmetrics are usually considered as the complement beyond citations. Altmetrics have some merits for evaluating. However, Altmetrics are only based on web usage statistics~\cite{barbaro2014altmetrics}. They are more easily manipulated by factitiously downloading, sharing, commenting, etc. What can be done to guarantee the credibility of data on social media for evaluating article impact? What can be measured by Altmetrics? How to select sources of data for Altmetrics? What relationships exist between Altmetrics and citations? Using data analysis techniques to explore Almetrics indicators in depth provides a possible solution to validating Altmetrics. There~are many explored opportunities in article impact researches.
\subsection{Benchmarks}
Available and credible benchmarks are key to measuring article impact. Despite past decades witnessed important progress, it is difficult to verify the performance of article impact metrics. Without right datasets and standards, developed metrics are not contextually robust and cannot be understood~\cite{wilsdon2015we}. Therefore, how to select benchmarks based on unified and consistent scholarly datasets with the aim of objectively quantifying impact is an important open issue.
\section{Open Issues and Challenges on Article Impact Prediction}
Despite our research has summarized article impact prediction so far, a great number of further issues and challenges call for our attention to predict impact  accurately. In this section, we point out some potential issues except for unified scholarly datasets and benchmarks.
\subsection{Sleeping Beauty}
Despite of the previously analyzed Sleeping Beauties phenomena, various issues remain to be addressed in the corresponding researches. How to identify Sleeping Beauties in science? How~to predict impact of Sleeping Beauties? Whether the trending topics are related to Sleep Beauties? Whether~the trending topics have contributed to predict Sleep Beauties? Whether the correlations between Sleep Beauties and different journals, between Sleep Beauties and institutions can influence the impact of Sleeping Beauties? Therefore, more efforts are needed to explore these critical scientific~problems.
\subsection{Multidimensional Prediction}
Despite pioneered researchers have obtained success from multidimensional perspective in predicting article impact, a full integration of multidimensional datasets needs to be explored in a further step. Characterizing the breadth and the depth of an article impact is unfortunately only from one single perspective. For example, previous researches generally focused on early citations to predict impact of an article~\cite{bruns2015research}. However, little attention has been paid to location information such as institutions and countries, social relationships and citation distributions for predicting impact. Therefore, future research needs to predict article impact from multiple dimensions.
\subsection{Rising Star}
Predicting the fast-rising citations for an article in the future provides valuable guidance to the academia. It can help the academia to find out popular topics or new topics, advanced techniques, significant findings, etc. Meanwhile, a direct benefit is to avoid wasting time in the ocean of scholarly data for researchers. What are the features contributed to enhance an article impact? Finding these features is beneficial to predict rising star in articles.

\section{Conclusions}
This article presents a detailed overview of evaluating and predicting article impact. It discusses the open issues and challenges that need to be solved in a further step. At first, we have given a simple introduction about article impact research. Next, we have elaborated on core methods and recent progress. Then, we have introduced some key techniques, and some opportunities can be seen by leveraging statistics, machine learning, data mining and network science techniques. Finally, we have presented open research issues regarding the assessment and prediction of article impact, and pointed out potential research directions.
%%%%%%%%%%%%%%%%%%%%%%%%%%%%%%%%%%%%%%%%%%
% Citations and References in Supplementary files are permitted provided that they also appear in the reference list here.

%%%%%%%%%%%%%%%%%%%%%%%%%%%%%%%%%%%%%%%%%%
\authorcontributions{X.B. conceived the study and wrote the manuscript; F.X. supervised the design and the development of the proposed study; F.Z. contributed statistical analysis work; X.B., H.L., F.Z., Z.N., X.K., I.L. and F.X. revised the manuscript; and all authors have read and approved the final manuscript.
} % please add author contributions

%%%%%%%%%%%%%%%%%%%%%%%%%%%%%%%%%%%%%%%%%%
\conflictsofinterest{The authors declare no conflict of interest.} % please add conflicts of interest

%=====================================
% References, variant A: internal bibliography
%=====================================

%\begin{thebibliography}{999}
%% Reference 1
%\bibitem[Author1(year)]{ref-journal}
%Author1, T. The title of the cited article. {\em Journal Abbreviation} {\bf 2008}, {\em 10}, 142-149.
%% Reference 2
%\bibitem[Author2(year)]{ref-book}
%Author2, L. The title of the cited contribution. In {\em The Book Title}; Editor1, F., Editor2, A., Eds.; Publishing House: City, Country, 2007; pp. 32-58.
%\end{thebibliography}

% The following MDPI journals use author-date citation: Arts, Econometrics, Economies, Genealogy, Humanities, IJFS, JRFM, Laws, Religions, Risks, Social Sciences. For those journals, please follow the formatting guidelines on http://www.mdpi.com/authors/references

%=====================================
% References, variant B: external bibliography
%=====================================
% \externalbibliography{yes}
% \bibliography{sample}

%%%%%%%%%%%%%%%%%%%%%%%%%%%%%%%%%%%%%%%%%%
%% optional
%\sampleavailability{Samples of the compounds ...... are available from the authors.}

%%%%%%%%%%%%%%%%%%%%%%%%%%%%%%%%%%%%%%%%%%
\end{document}